# Models as Values in a Model Expression Algebra:
# A Functional Approach to Model Driven Engineering

Siniša Nešković, IT University of Copenhagen, Copenhagen, Denmark,
sinn@itu.dk, 0000-0002-8309-5847

Dejan Stojimirović, University of Belgrade, Faculty of Organizational Sciences, Belgrade, Serbia,
dejan.stojimirovic@fon.bg.ac.rs, 0000-0003-0444-295X

**Abstract**

This paper proposes a functional foundation for model driven engineering that unifies model construction, metamodels, templates, and transformations under a single formalism: the model expression algebra. In this algebra, models are values, model expressions are terms, and evaluation is the interpretation homomorphism from terms to values. Model expressions are composed from four operators: model creation and element creation operators, reference operators for retrieving models and elements, and computation operators that embed functional computations. Metamodels are type schemas that constrain the algebra, and model templates, understood as parameterized model expressions, are formalized as open terms with free variables. Model transformations then arise naturally as model templates whose input parameter is a source model. We prove type preservation under evaluation and type safety of transformation execution. Since models are themselves model elements, the algebra also supports megamodels and weaving models without additional mechanisms. The approach is realized through an embedded domain-specific language (DSL) that demonstrates how a single mainstream language can serve simultaneously as the metamodeling, model construction, and transformation language, with formal guarantees enforced by the language's type system.

## 1. Introduction

Model Driven Development (MDD) [1], [2] is an approach to software development which promotes the idea that the primary result of software development is not executable source code, but the knowledge about a developed system expressed through models. Furthermore, MDD seeks to automate software development through systematic usage of model transformations, which define how high-level models are mapped to other design-level models or to final implementation artifacts such as executable source code. The OMG Model Driven Architecture (MDA) [3] and the Eclipse Modeling Framework (EMF) [4] are the most prominent realizations of this vision, forming the core of a mature model driven engineering (MDE) ecosystem of standards, languages, and tools.

Yet after more than two decades of research and industrial use, MDD has not achieved the widespread adoption that its proponents envisioned [5]. Several persistent challenges contribute to this. Models tend to drift out of sync with implementation during maintenance, limiting their usefulness beyond the early stages of development [6].Transformation languages such as QVT [7] and ATL [8] require separate tools and specialized expertise, creating adoption barriers. The mathematical foundations of MDE remain fragmented, with no unified formal framework that covers models, metamodels, and transformations in a compositionally coherent manner. Most critically, current MDE approaches offer surprisingly limited mechanisms for what should be a fundamental capability: the systematic customization, parameterization, and reuse of models and transformations.

We argue that several of these limitations stem from the dominant object-oriented (OO) paradigm underlying current MDE. In mainstream OO frameworks that underpin MDA and EMF, a model is a collection of interconnected objects instantiated from metamodel classes. This representation treats models as fixed, passive data structures. Once constructed, a model is that particular model. If a practitioner needs a family of related models, a parameterized model variant, or a reusable and customizable transformation rule, there is no natural mechanism to express any of these within the OO paradigm. Existing reuse mechanisms for models (UML templates, package templates) and for transformations (rule inheritance, rule composition) have been developed separately, and as Kusel et al. have shown [9], none fully addresses reuse across metamodel boundaries.

This paper proposes an alternative foundation for MDE, rooted in the functional programming paradigm, that addresses these limitations through a single formalism: the model expression algebra. The key idea is that models are values in this algebra, and a model expression is a term-algebraic composition of four operators:

a) model creation operators ($\mu$) that create models as typed containers with their own type schema and namespace for contained elements,

b) element creation operators ($\varepsilon$) that introduce typed model elements with properties,

c) reference operators ($\rho$) that resolve named elements to enable cross-references in the evaluated model value, and

d) computation operators ($\kappa$) that embed functional computations producing sub-expressions within the algebra.

This algebraic perspective reveals that the relationship between what modelers write (expressions) and what models are (values) is precisely the relationship between terms and their interpretations in universal algebra. The evaluation function serves as the interpretation homomorphism.

This shift from models-as-objects to models-as-algebraic-terms brings the expressive mechanisms of functional programming to model-driven engineering. Parameterization becomes natural, since a model expression with free variables is a template, an open term in the algebra, that defines an entire family of models from a single definition. Composition is inherent, as element expressions nest hierarchically, computation operators generate sub-expressions dynamically, and the algebra provides compositional semantics for building complex models from simpler parts. Model transformations are then simply templates whose input parameter is a source model, producing a target model as output. Metamodels become type schemas that constrain the algebra, specifying which constructors exist and what their argument types must be.

The formalization deliberately addresses only what is genuinely specific to models and model transformations. Other useful capabilities needed in model transformations, such as aspect-oriented composition, packages, import/export, and persistence, belong to general-purpose programming and are already well defined and understood.

This separation of concerns principle also drives our implementation choice. Rather than extending existing specialized MDE frameworks such as EMF, ATL, or QVT, or building a new dedicated transformation language, we realize the approach within a mainstream programming ecosystem, reusing its existing and familiar concepts, languages and tools directly. Synthesis, an embedded DSL in TypeScript using JSX as concrete syntax for model expressions, developed within the AelasticS open-source project [10], is the practical realization of this choice [11].

Based on these ideas, the paper makes four contributions:

(1) A formal model expression algebra with evaluation semantics, compositional construction, and metamodels as type schemas.

(2) A formalization of model templates as parameterized model expressions, which enhance existing customization and reuse mechanisms in MDE.

(3) A formalization of model transformations as templates over source models, unifying model construction and model transformation within a single algebraic framework, with proven type safety and type preservation.

(4) A realization of our approach through an embedded DSL demonstrating that it is practically implementable.

The remainder of this paper is organized as follows. Section 2 motivates the functional paradigm through a running example. Section 3 establishes the formal framework. Section 4 describes the Synthesis realization and demonstrates a complete end-to-end transformation. Section 5 discusses related work, and Section 6 concludes.

## 2. Motivating Example

To make the ideas introduced above concrete before the formal treatment, we show how a model can be viewed as a hierarchical composition of element creation functions and how parameterization naturally yields model families.

Consider an organization that uses approval workflows for documents. A simple workflow consists of a Write task followed by two parallel Approve tasks. In the functional view, this workflow is expressed as a hierarchical composition of element creation functions:

```
me1 = () =>
  Process("Approval",
    Sequence(
      Task("Write"),
      Parallel(
        Task("Approve 1"),
        Task("Approve 2")
      )
    )
  )
```

Each name, Process, Sequence, Task, Parallel, denotes an element creation operator that creates a model element of the corresponding type and connects it to the elements created by its nested calls. The body of me1 is a model expression, a term in the model expression algebra. Each element creation operator (ε) is a constructor node in the expression tree. Calling me1() evaluates the expression and produces a model value, the result of creating each element and assembling them into an interconnected structure.

In this simple case, every position in the expression tree is occupied by an element creation operator. However, any position in the tree that can hold an element creation operator can instead hold a computation operator (κ), a functional computation that produces element creation operators. An operator κ can map over a collection to generate a variable number of element creation operators, select between alternative structures based on a condition, or call other functions that return operators. This is what makes parameterization possible.

Consider a more realistic requirement. Different departments need different approval workflows (business processes), varying in the number of approvers and whether they work in parallel or sequentially. Adding parameters to the function turns the closed model expression into an open one, a model template with free variables:

```
me2 = (doc, approvers, mode) =>
  Process(`${doc} Approval`,
    Sequence(
      Task(`Write ${doc}`),
      mode === "parallel"
        ? Parallel(
            map(1.. approvers, i => Task(`Approval of ${doc} - stage ${i}`))
          )
        : Sequence(
            map(1.. approvers, i => Task(`Approval of ${doc} - stage ${i}`))
          )
    )
  )
```

me2("Contract", 3, "parallel") produces a process with three parallel approval tasks. me2("Invoice", 2, "sequential") produces a process with two sequential approval tasks, a structurally different model from the same definition. The two results differ in cardinality (three versus two approval tasks), topology (parallel versus sequential), and element properties (process and task names derived from the document name parameter). The map() function and the conditional are computation operators (κ) embedded within the model expression. They produce sub-expressions that create elements of the types expected at those positions. Thus, calling the same model expression with different arguments can generate an entire family of structurally different models.

Calling me2 with concrete arguments reduces the κ nodes, and the result is a pure model expression, a ground term containing only element creation operators. Evaluating this ground term then creates each element and assembles them into a model. We formalize this precisely in Section 3.

## 3. Formal Framework

The proposed formalization is representation-agnostic, defining semantic requirements using sets, functions, and tuples without prescribing how these are encoded in a particular programming language. Section 3.1 defines model elements, models, and the operations by which models are constructed. Section 3.2 defines the type constructor system (M3 level in MDA). Section 3.3 formalizes metamodels as type schemas (M2 level). Section 3.4 introduces model expressions and templates. Sections 3.5 and 3.6 formalize model transformations and their execution semantics.

### 3.1. Models and Model Elements

**Definition 1 (Model Element).** A model element x ∈ E (where E is the set of all model elements) is characterized by:

(i) (i) a qualified name qname(x) ∈ QName,

(ii) a type $\tau(x)$ ∈ T where T is a set of types (formalized in Section 3.2),

(iii) a property record props(x), a finite set of name–value pairs assigning values to the properties of x. The type and multiplicity of each property are determined by $\tau(x)$ (formalized in Sections 3.2–3.3). Property values may be data values (for base-typed properties), references to model elements (for entity-typed properties), sequences (for array-typed properties), or absent (for optional-typed properties).

**Notation (Qualified Name).** A qualified name (QName) is a path expression using "/" as delimiter. It consists of a sequence of name segments $n_1/n_2/\ldots/n_k$ where k ≥ 1. A QName may be prefixed with "/" for an absolute path, or with one or more "../" segments for a relative path. A QName with a single segment and no prefix is a simple name. The final segment is the local name.

**Definition 2 (Model).** A model m is a model element (Definition 1) with a type $\tau(m)$ and a property record props(m), together with:

(i) a type schema $\Sigma(m)$ (formalized in Section 3.3) that constrains which types, properties, and multiplicities are valid for elements created within m, and

(ii) an element collection E(m) ⊆ E, the set of model elements belonging to m, where name(x) is unique for every x ∈ E(m).

**Axiom 1 (Root Model).** A root model $m_0$ exists with $E(m_0) = \emptyset$ and a universal type schema $\Sigma_0 = (T, \tau_0)$ where T is the full type universe (Definition 8) and $\tau_0$ is the base model type. Since every type schema $\Sigma = (T_\Sigma, \tau_{model})$ satisfies $T_\Sigma \subseteq T$ and $\tau_{model} \sqsubseteq \tau_0$, any element may be added to $E(m_0)$. It serves as the outermost namespace for all element and model constructions (Definitions 3 and 4).

**Definition 3 (Element Creation).** The element creation function

$$newElement: M \times Name \times T \times PR \to M \times E$$

takes a model m, a local name n, an entity type t from the model's type schema, and a property record pr. It produces a pair (m′, x) where:

(i) x is a new element with $qname(x) = qname(m)/n$, $\tau(x) = t$, and $props(x) = pr$,

(ii) $E(m') = E(m) \cup \{x\}$, and

(iii) no other element in E(m) has local name n.

Property values in pr that are entity-typed hold references to model elements produced by prior invocations of *newElement* or retrieved by the reference function (Definition 6).

**Definition 4 (Model Creation).** The model creation function

$$newModel: M \times Name \times \Sigma \times PR \to M \times M$$

takes a model m, a local name n, a type schema Σ (Definition 11), and a property record pr. It proceeds in two steps:

(a) It invokes $newElement(m, n, \tau_{model}, pr)$ (Definition 3), where $\tau_{model}$ is the model type of Σ, to produce (m′, x) where x is added to E(m).
(b) It initializes x as a model by setting $E(x) = \emptyset$ and binding Σ to x, constraining which types, properties, and multiplicities are valid for elements subsequently created within x.

The result is (m′, x) where x is both an element of m′ and an empty model ready to receive its own elements.

**Definition 5 (Element Update).** The element update function

$$updateElement : M \times E \times PR \to M \times E$$

takes a model m, an existing element $x \in E(m)$, and additional properties pr. It produces (m′, x′) where:

(i) $props(x') = props(x) \oplus pr$, where pr overrides matching property names and adds new ones. For array-typed properties, the new value is concatenated to the existing sequence.
(ii) m′ is m with x replaced by x′.

The type and name of the element are preserved. Multiplicity constraints are not enforced by updateElement. They are verified as part of model conformance (Definition 12).

**Definition 6 (Element Reference).** The element reference function

ref : M × QName ⇀ E

is a partial function that, given a model m and a qualified name q:

(i) If q is a simple name n, returns the element x ∈ E(m) where the local name of qname(x) = n.
(ii) If q is a relative path, navigates from m upward through parent models for each "../" prefix, then downward through named sub-models for each subsequent segment, and returns the element matching the final segment.
(iii) If q is an absolute path, navigates from the root model $m_0$ (Axiom 1) downward through named sub-models for each segment, and returns the element matching the final segment.

No new element is created and no model state changes. If any segment along the path does not resolve, construction halts with an error.

**Definition 7 (Well-Formedness).** A model m is well-formed if and only if:

(i) every element in E(m) has a unique name within the model's namespace,

(ii) for every element x ∈ E(m) and every property p in props(x), the value conforms to typeOf(τ(x), p) and satisfies mult(τ(x), p) (formalized in Definition 12), and

(iii) every name used in a ρ invocation during construction resolves to an existing element at the time of invocation.

Conditions (i) and (ii) are statically enforceable when the type schema is realized as a type system in the implementation language (Section 4). Condition (iii) requires runtime verification.

### 3.2. Type Constructors: The Meta-Meta Model

In the OMG MDA four-level architecture, the meta-meta level (M3) defines the constructs from which metamodels are built. In our functional approach, the M3 level is realized as a system of type constructors, functions that produce types. Metamodels are assembled by applying these constructors (Section 3.3). The sets T (types) and PR (property records) introduced abstractly in Definitions 1 and 3 are grounded by this system. A property specification PS is a finite partial function from property names to type–multiplicity pairs (formalized in Definition 9).

**Definition 8 (Type Constructor System).** A type constructor system (meta-meta model) is a tuple $\Gamma = (T_B, C, T, \sqsubseteq)$ where:

- $T_B$ is a finite, non-empty set of base types (primitive types), including at minimum string, number, and boolean.
- C is a finite set of type constructors, each a function that takes types and auxiliary parameters as arguments and produces a type.
- $T \supseteq T_B$ is the type universe, the smallest set containing $T_B$ and closed under all constructors in C.
- $\sqsubseteq \subseteq T \times T$ is a subtype relation on T, a partial order (Definition 10).

The core constructors in C are:

- Entity type constructor: c_ent : Name × PS → T. Produces an identifiable type whose instances are model elements (Definition 1) with unique names managed by the model (Definition 2).
- Subtype constructor: c_sub : T × Name × PS → T. Produces a type that specializes $\tau\_p$ by adding properties ps_e to those inherited from $\tau\_p$. The resulting type satisfies c_sub($\tau\_p$, n, ps_e) ⊑ $\tau\_p$.
- Array type constructor: c_arr : T → T. Produces the type of ordered, variable-length sequences.
- Optional type constructor: c_opt : T → T. Produces a type whose instances are either a value of type $\tau$ or absent.
- Reference type constructor: c_ref : T → T. Takes a type name as a string and produces the corresponding type, enabling recursive type definitions and forward references.
- Instance reference type constructor: c_iref : Σ × T_E → T. Takes a type schema $\Sigma'$ and an entity type $\tau \in T\_E(\Sigma')$ and produces a type whose values are qualified names (QName) designating instances of type $\tau$ in models conforming to $\Sigma'$.

The set C is intentionally minimal but extensible. Specific implementations may introduce additional constructors (e.g., union types, intersection types, set types, map types, enumeration types) without affecting the framework's core properties.

**Definition 9 (Property Specification).** A property specification is a finite partial function ps : PropName ⇀ T × Mult where PropName is a set of property names and Mult = {1, 0..1, 0.., 1..} is a set of multiplicity indicators. For a type $\tau$ with property specification ps, we define props($\tau$) = dom(ps), typeOf($\tau$, p) = $\pi_1$(ps(p)), and mult($\tau$, p) = $\pi_2$(ps(p)). Property types may be base types, entity types, or types produced by c_arr, c_opt, c_ref, or c_iref.

**Definition 10 (Subtype Relation).** The subtype relation ⊑ on T is the smallest partial order satisfying: (i) Reflexivity: $\tau \sqsubseteq \tau$ for all $\tau \in T$. (ii) Transitivity: if $\tau_1 \sqsubseteq \tau_2$ and $\tau_2 \sqsubseteq \tau_3$, then $\tau_1 \sqsubseteq \tau_3$. (iii) Antisymmetry: if $\tau_1 \sqsubseteq \tau_2$ and $\tau_2 \sqsubseteq \tau_1$, then $\tau_1 = \tau_2$. (iv) Subtype construction: $\kappa_{sub}(\tau_p, n, ps_e) \sqsubseteq \tau_p$.

Since c_sub adds ps_e to the parent's property specification without modifying inherited properties, a subtype inherits all properties of its parent with identical types and multiplicities. By transitivity of ⊑, inheritance is transitive: if $\tau\_s \sqsubseteq \tau\_p \sqsubseteq \tau\_a$, then $\tau\_s$ inherits the properties of $\tau\_a$ as well.

### 3.3. Metamodels as Type Schemas

A metamodel is formalized as a type schema, a finite collection of types drawn from the type universe T (Definition 8) that together define a modeling language.

**Definition 11 (Type Schema).** A type schema over a type constructor system $\Gamma = (T_B, K, T, \sqsubseteq)$ is a tuple $\Sigma = (T_\Sigma, \tau_{model})$ where: $T_\Sigma \subseteq T$ is a finite, non-empty set of schema types, and $\tau_{model} \in T_\Sigma$ is the entity type assigned to models conforming to $\Sigma$. The schema types are partitioned into: $T_E \subseteq T_\Sigma$, the set of entity types (types produced by $\kappa_{ent}$ or by $\kappa_{sub}$ applied to an entity type), and $T_V = T_\Sigma \setminus T_E$, the set of value types. We write $T_E(\Sigma)$ for the set of entity types of schema $\Sigma$, and $T_V(\Sigma)$ for its value types. The type schema must satisfy the following conditions:

- *Type closure*: for every entity type $\tau \in T_E$ and every property $p \in props(\tau)$, $typeOf(\tau, p)$ is in $T_\Sigma \cup T_B$, and every entity type referenced within $typeOf(\tau, p)$ is in $T_E$.
- *Subtype closure*: for every entity type $\tau\_s \in T_E$ produced by $c\_sub(\tau\_p, n, ps\_e)$, the parent type $\tau\_p \in T_E$.

**Definition 12 (Model Conformance).** A model m (Definition 2) conforms to a type schema $\Sigma = (T_\Sigma, \tau_{model})$, written m : $\Sigma$, if and only if:

(i) Model typing: $\tau(m) \sqsubseteq \tau_{model}$.

(ii) (ii) Element typing: for every element $x \in E(m)$, $\tau(x) \in T_E$.

(iii) (iii) Property conformance: for every element $x \in E(m)$ and every property $p \in props(\tau(x))$, the value of p in $props(x)$ conforms to $typeOf(\tau(x), p)$ and satisfies $mult(\tau(x), p)$.

**Proposition 1 (Operational Type Safety).** Let m be a model constructed through a sequence of element creation (Definition 4) and element *updateElement* (Definition 5) functions against an initially empty model $m_0 = newModel(\tau_{model})$. If every element creation invocation uses a type from $T_E$ with a conforming property record, then the resulting model m conforms to $\Sigma$.

*Proof.* Model typing holds because *newModel* assigns $\tau_{model}$. Element typing holds because each element creation uses a type from $T_E$, and elements are only added, never removed. Property conformance holds because each operation individually satisfies the constraints of Definition 12, and subsequent updateElement operations only add or override property values that conform to the type schema (*updateElement* only merges additional data). Therefore, all conformance conditions of Definition 12 are satisfied in the final model. □

### 3.4. Model Expressions and Templates

This section presents the algebra for compositional model construction. A model expression is a term in a term algebra whose evaluation produces model elements via the primitives of Section 3.1. The algebra has four operators: the model creation operator $\mu$ (which invokes *newModel* to create a model with its own type schema and namespace), the element creation operator $\varepsilon$ (which invokes *newElement* to create an element), the reference operator $\rho$ (which invokes ref to retrieve existing elements), and the computation operator $\kappa$ (which computes sub-expressions from parameters)

#### 3.4.1. Model Expressions and Evaluations

**Definition 13 (Model Expression).** Let $\Sigma = (T_\Sigma, \tau model)$ be a type schema (Definition 11) and $\Theta$ a parameter space. A model expression over $\Sigma$ and $\Theta$ is an inductively defined term built from four operators:

(i) **Model creation operator.** $\mu(n, \Sigma', p, (a_1, cs_1), \ldots, (a_m, cs_m))$ where $n \in$ Name is the local name, $\Sigma'$ is a type schema, $p \in P$ is a property record, and each $(a_i, cs_i)$ pair specifies an entity-typed property and its child expressions.

(ii) **Element creation operator.** $\varepsilon(n, t, p, (a_1, cs_1), \ldots, (a_m, cs_m))$ where $n \in$ Name is the local name, $t \in T\_\Sigma$ is an entity type, $p \in P$ is a property record for the non-entity-typed properties of $t$, each $a_i$ is an entity-typed property of $t$, and each $cs_i$ is a finite sequence of model expressions, the child expressions for entity-typed property $a_i$.

(iii) **Reference operator.** $\rho(n)$ where $n \in$ QName is a qualified name.

(iv) **Computation operator.** $\kappa(f)$ where $f : \Theta \rightarrow$ Expr* is a computable function that, given parameter values $\theta \in \Theta$, produces a finite sequence of model expressions. Computation operators introduce programmatic structure, such as iteration and conditionals, into model construction.

A model expression containing only $\mu$, $\varepsilon$, and $\rho$ operators is a pure (or ground) expression. An expression containing $\kappa$ operators whose functions reference parameters from a non-empty $\Theta$ is a template (Section 3.4.3), an open term whose evaluation depends on parameter bindings. When discussing the structure of a model expression as a tree, we also refer to its constituent operators as nodes.

The semantics of these four operators is given by the evaluation function, which maps a model expression to elements (including models) within their enclosing models. Evaluation proceeds in document order, a top-down, left-to-right, depth-first traversal of the expression tree.

**Definition 14 (Evaluation).** The evaluation function

$$\text{eval} : \text{Expr} \times \Theta \times M \rightarrow M \times E^*$$

maps a model expression, parameter values, and an enclosing model to an updated model and a sequence of elements. Evaluation is defined inductively on the structure of the expression.

**Model creation operator.** For $\mu(n, \Sigma', p, (a_1, cs_1), \ldots, (a_m, cs_m))$, *newModel(m, n, $\Sigma'$, p)* (Definition 4) is invoked to produce a new model $m'$ within the enclosing model $m$. Each child expression sequence $cs_i$ is then evaluated left-to-right with $m'$ as the enclosing model. Elements produced by child expressions are added to $E(m')$. The *updateElement* function (Definition 5) sets the entity-typed properties of $m'$ to the resulting child elements. Because $m'$ exists before its children are evaluated, reference operators in children can resolve to $m'$ or to preceding siblings.

**Element creation operator.** For $\varepsilon(n, t, p, (a_1, cs_1), \ldots, (a_m, cs_m))$ *newElement(m, n, t, p)* (Definition 3) is invoked to create the element, initially without entity-typed properties. Child expressions are evaluated left-to-right, and *updateElement* (Definition 5) connects them to the parent. Because the parent exists before its children are evaluated, reference operators in children can resolve to the parent or to preceding siblings.

**Reference operator.** For $\rho(q)$, the partial function ref(m, q) (Definition 6) is evaluated. If q resolves, return (m, [ref(m, q)]). Otherwise, construction halts with an error. Since a model is a model element (Definition 2), ref may return a model, which can then be connected to a parent via an entity-typed property, just as any other element.

**Computation operator.** For κ(f), the function f is applied to θ to produce sub-expressions, which are then recursively evaluated.

**Proposition 2 (Evaluation Termination).** Evaluation of a model expression terminates provided that:

(i) the expression tree is finite,

(ii) every computation function f terminates for the given θ, and

(iii) every computation function produces a finite sequence of sub-expressions whose maximum nesting depth of κ nodes is bounded.

*Proof.* Pure expressions are finite trees, and each node is visited exactly once in the depth-first traversal. For expressions containing κ nodes, each reduction replaces one κ node with a finite sequence of sub-expressions. Conditions (i) and (iii) together bound the total number of reductions. Since each reduction terminates (condition ii) and produces finitely many sub-expressions, the overall evaluation terminates. □

**Theorem 1 (Type Preservation under Evaluation).** Let expr be a model expression over type schema $\Sigma$ such that every ε node ε(t, p, …) satisfies $t \in T\Sigma$ with a conforming property record, every μ node μ($\Sigma'$, p, …) specifies a valid type schema, and every computation function f in every κ node produces well-typed sub-expressions for the given θ. Then for any model $m : \Sigma$, if eval(expr, θ, m) = (m′, es), then $m' : \Sigma$.

*Proof.* By structural induction on the expression. For μ nodes, *newModel* (Definition 4) invokes *newElement* (Definition 3) to add a model element of type $\tau\Sigma'$ to the enclosing model, preserving conformance of the enclosing model by Proposition 1. The child expressions are evaluated within the new model m′ constrained by $\Sigma'$, and by the induction hypothesis each child preserves conformance of m′ with respect to $\Sigma'$. For ε nodes, the element creation *newElement* (Definition 3) adds an element of type $t \in T\Sigma$ with conforming properties, preserving conformance by Proposition 1. The subsequent *updateElement* calls (Definition 5) set entity-typed properties whose types are constrained by the type schema, and *update* preserves type and name, so conformance is maintained. For ρ nodes, the model is unchanged, trivially preserving conformance. For κ nodes, f(θ) produces well-typed sub-expressions by assumption, and by the induction hypothesis each sub-expression's evaluation preserves conformance. Since the enclosing model is passed to all evaluations, each starting from a conforming state, the final model m′ conforms to $\Sigma$. □

**Remark (Reference Safety).** Evaluation succeeds without construction errors if and only if, at each point during evaluation a ρ(q) node is evaluated, the qualified name q resolves in the enclosing model. This condition is naturally satisfied in document-order evaluation when the referenced element appears earlier in the expression.

These definitions and results can be interpreted in the framework of universal algebra. The model creation operator μ and element creation operator ε are constructors, operations that build new values (models and model elements) from components. The computation operator κ is a derived operation whose functions produce sequences of model expressions. The reference operator ρ is a lookup that evaluates to existing elements in the model without constructing new ones. It does not correspond to any standard construct in classical term algebras, where all values are produced by constructors.

### 3.4.2. Model Templates

**Definition 15 (Model Template).** A model template over type schema $\Sigma$ and parameter space $\Theta$ is a model expression containing computation operators $\kappa$ whose functions may reference parameters from $\Theta$. The template defines a family of models: $F(expr) = \{ eval(expr, \theta, m_0) \mid \theta \in \Theta \}$. Different values of $\theta$ may produce different numbers of elements, different connection topologies, and different property values.

Three special cases are worth distinguishing. When $\Theta = \emptyset$ and the expression contains no $\kappa$ nodes, it is a model literal (a ground term). When $\Theta = \emptyset$ but $\kappa$ nodes are present, the $\kappa$ functions reference no external parameters, and the expression produces the same elements on every evaluation. When $\Theta \neq \emptyset$, the expression is a template proper, a parameterized model expression whose structure varies with parameter values. Since each instantiation is an evaluation of a model expression, type preservation (Theorem 1) applies to every member of the family $F(expr)$, provided that every computation function produces well-typed sub-expressions for all $\theta \in \Theta$.

**Definition 16 (Higher-Order Template).** A higher-order model template is a model template whose parameter space $\Theta$ contains at least one template-typed parameter, a parameter whose value is itself a function $\Theta' \to Expr$ producing model expressions.

Since $\kappa$ operators accept arbitrary computable functions and templates are themselves functions from parameters to expressions, higher-order templates require no additional formal apparatus beyond Definition 13. Section 4.3 demonstrates higher-order templates concretely.

### 3.4.3 Megamodels

Since a model is a model element (Definition 2) and its type $\tau\_model$ is an entity type in the enclosing model's type schema, a $\mu$ node may appear as a child of another $\mu$ node. The inner model is both an element in the enclosing model's element collection and an independent namespace with its own element collection and type schema. This enables compositional nesting of models within models.

Different sub-models within a composite structure may conform to different type schemas, since each $\mu$ node binds its own $\Sigma$ independently. The only constraint is that the inner model's type must appear in the enclosing model's type schema, as required by Definition 3 (element creation). The type closure condition of Definition 11 (Type Schema) ensures that all referenced types are present.

This gives a formal basis for megamodels, a concept introduced by Bézivin et al. [10] as a model whose elements are themselves models. Bézivin et al. [11] further distinguish modeling-in-the-small (constructing individual models) from modeling-in-the-large (managing collections of interrelated models). A megamodel is a model expression whose outermost $\mu$ contains inner $\mu$ nodes as children. No additional definition or mechanism is needed. The typing, conformance, and evaluation semantics of Section 3.4.1 apply directly.

A type schema that includes model types among its entity types naturally serves as a megamodel schema. Since models are model elements, they participate in entity-typed properties just as any other element. Entity-typed properties in such a schema can connect models to models, models to elements, and elements to elements.

Existing models can also be composed into megamodels via reference expressions. A ρ node may return a previously created model, since ref (Definition 6) operates on model elements and every model is a model element. The retrieved model is then connected to its parent via an entity-typed property, just as any other child element.

The root model $m_0$ (Axiom 1) can be understood as the outermost megamodel. Its universal type schema admits any model type, so all top-level models are elements of $E(m_0)$.

### 3.4.4 Weaving Models

The instance reference type constructor c_iref (Definition 8) enables entity types whose properties refer to elements in other models. A c_iref property stores a qualified name as data, designating an element of a specified type in a model conforming to a specified schema. The stored value is the reference itself, not the referenced element. This is analogous to a foreign key in the relational model. The correspondence is recorded as data, and dereferencing is a separate operation via ref (Definition 6).

A type schema whose entity types consist primarily of c_iref properties is a weaving schema. Models conforming to such a schema are weaving models, first-class models that capture typed links between elements of distinct models. The concept of model weaving was introduced by Del Fabro et al. [12]. In our framework, weaving models require no dedicated mechanism. A weaving model is constructed via μ and ε operators, conforms to its own type schema, and participates in the evaluation and conformance semantics of Sections 3.4.1 and 3.3 without special cases.

Property conformance (Definition 12) for c_iref values requires only that the stored value is a valid qualified name. The referenced element need not exist at construction time. This means a weaving model can be constructed, stored, and validated independently of the models it references. Referential integrity, verifying that all stored references resolve against a given set of models, is an optional cross-model validation, not a precondition for well-formedness.

Weaving models support any scenario requiring explicit correspondences between elements of independently defined models, such as transformation traceability (Section 3.6), model integration, or model comparison.

## 3.5. Transformation Specifications

A transformation is a collection of typed rules, functions that map source model elements to target model expressions. Each rule is a model template (Definition 15) whose parameter space consists of source model elements.

**Definition 17 (Transformation Rule).** A transformation rule is a typed function $r : E_\tau \to \text{Expr}(\Sigma_T)$ where $E_\tau = \{x \in E(m_S) \mid \tau(x) \sqsubseteq \tau_S\}$ is the set of source elements of type $\tau_S$ or any subtype, $\tau_S \in TE(\Sigma_S)$ is the source entity type, $\Sigma_T$ is the target type schema, and $\text{Expr}(\Sigma_T)$ denotes a model expression over ΣT (Definition 13). Each rule takes a source element from Eτ and produces a target model expression.

Because each rule produces a model expression, it inherits the full expressive power of the algebra, including computation operators κ for iterative and conditional generation.

**Definition 18 (Transformation Specification).** A transformation specification is a tuple TSpec = ($\Sigma_S$, $\Sigma_T$, R, SP, SO, dispatch, $r_0$) where: $\Sigma_S$ and $\Sigma_T$ are the source and target type schemas, R is a finite set of rules, SP ⊆ R is a set of specialization points, SO is a finite set of specialization options, dispatch : $T_E(\Sigma_S)$ → R maps each source entity type to a rule, and $r_0 \in$ R is the top-level rule.

**Definition 19 (Rule Dispatch).** Given a source element e ∈ E($m_S$) with type τ(e), the dispatch function selects the most specific applicable rule. We write $\tau_S$(r) for the source type of rule r (Definition 17). Then dispatch(τ(e)) = the unique rule r ∈ R ∪ SO such that τ(e) ⊑ $\tau_S$(r), and there is no other applicable rule r′ with τ(e) ⊑ $\tau_S$(r′) and $\tau_S$(r′) ⊑ $\tau_S$(r). Specialization options take priority over their parent specialization points for matching subtypes.

**Proposition 3 (Dispatch Determinism).** If for every specialization point sp ∈ SP, the source types of its options are pairwise incomparable under ⊑, then dispatch is deterministic. For every source element, exactly one rule is selected.

Consider a source element e with type τ(e). At least one rule is applicable, because the specialization point itself serves as fallback. Suppose two distinct options $so_1$, $so_2$ ∈ options(sp) both apply. Then τ(e) ⊑ $\tau_S(so_1)$ and τ(e) ⊑ $\tau_S(so_2)$. Pairwise incomparability means neither $\tau_S(so_1)$ ⊑ $\tau_S(so_2)$ nor the reverse, so the most-specific criterion cannot prefer one over the other, contradicting uniqueness. Therefore at most one option applies, and combined with the fallback guarantee, exactly one rule is selected. □

**Definition 20 (Dispatch Completeness).** A transformation specification TSpec is complete if for every entity type τ ∈ $T_E(\Sigma_S)$, there exists a rule r ∈ R such that τ ⊑ $\tau_S$(r).

### 3.6. Transformation Execution Semantics

**Definition 21 (Transformation Execution).** Given a transformation specification TSpec = ($\Sigma_S$, $\Sigma_T$, R, SP, SO, dispatch, $r_0$) and a source model $m_S$ conforming to $\Sigma_S$, together with an initial target model $m_{0T}$ = newModel($\tau_{modelT}$), the execution of TSpec on $m_S$ is: exec(TSpec, $m_S$) = eval($r_0$(root($m_S$)), ∅, $m_{0T}$).

Each rule r, when invoked on a source element e, produces a target model expression whose evaluation creates elements in the target model. The target model accumulates elements monotonically, so elements created by earlier rule executions are visible to later executions via ref (Definition 6).

**Definition 22 (Execution Trace).** An execution trace is a sequence Tr = ⟨$tr_1$, …, $tr_n$⟩ where each entry $tr_k$ = ($e_k$, $r_k$, $t_k$) records the source element being transformed, the dispatched rule, and the set of target elements produced. The trace provides bidirectional traceability.

**Theorem 2 (Type Safety of Execution).** If every rule in R ∪ SO produces a well-typed model expression over $\Sigma_T$ (in the sense of Theorem 1) and the source model $m_S$ conforms to $\Sigma_S$, then the target model $m_T$ produced by exec(TSpec, $m_S$) conforms to $\Sigma_T$.

*Proof.* By structural induction on the rule invocation tree. Base case: a rule that makes no further invocations produces a target model expression whose evaluation, by Theorem 1, yields a conforming model. Inductive step: a rule that invokes other rules on related source elements receives a conforming model state after each invocation (by the induction hypothesis). Since the rule's own expression is well-typed and evaluation starts from a conforming state, Theorem 1 guarantees that the resulting model state is also conforming. The root rule $r_0$ composes all invocations, and since each preserves conformance, the final model $m_T$ conforms to $\Sigma_T$. □

**Proposition 4 (Trace Completeness and Consistency).** For a transformation execution producing trace Tr and target model $m_T$: (i) every target element t ∈ E($m_T$) appears in at least one trace entry, (ii) every trace entry records the rule selected by dispatch, and (iii) every source element reached through rule invocations from $r_0$ has at least one trace entry.

*Proof.* Every target element is produced by evaluating some rule's model expression, and every rule evaluation generates a trace entry, establishing (i). The execution semantics (Definition 21) require that every invocation passes through dispatch, establishing (ii). A trace entry is created for every rule invocation, establishing (iii). □

This completes the formal framework. The foundation comprises four parts: the operational part (Section 3.1) defines models and primitives, the type system (Sections 3.2–3.3) constrains construction through type constructors and type schemas, the algebraic part (Section 3.4) introduces model expressions as a term algebra with evaluation semantics and templates, and the transformation part (Sections 3.5–3.6) formalizes typed rules with polymorphic dispatch and execution semantics. The key feature is that transformation rules are model templates parameterized by source elements, and the same algebraic framework governs both model construction and model transformation.

## 4. Realization in the Synthesis

To demonstrate that the formal framework of Section 3 is practically realizable, we have implemented a free, open-source embedded domain-specific language (eDSL) in TypeScript and JSX notation, named Synthesis. The source code is available on GitHub [10], and the project website [11] provides documentation and a tool for executing transformations. An eDSL reuses the syntax, compiler, and type system of its host language while providing domain-specific abstractions as library constructs rather than new syntax. Synthesis follows this pattern precisely. Every formal concept from Section 3, from model elements and type schemas through the model expression algebra to transformation specifications, maps to a specific TypeScript language mechanism.

Synthesis builds on two supporting packages that are general-purpose:

- The **aelastics-types** package provides functions corresponding to the type constructors of Definition 8, from which metamodels are assembled as runtime type descriptors.

- The **aelastics-store** package is a generic object store that creates and manages object instances of types defined by aelastics-types, maintains collections, and supports key-based lookup.

Synthesis adds model awareness on top of these packages. It defines ModelElement (Definition 1) and Model (Definition 2) as types using aelastics-types, designating the element name as the identifying key. It introduces ModelStore, which plays the role of the root model $m_0$ (Axiom 1), imposing name uniqueness within model namespaces and providing the model-level operations of Definitions 3, 4, and 6. The evaluation of model expressions (Definition 14) is realized through JSX rendering, as described in Section 4.3. Synthesis also provides the transformation engine with decorator-based polymorphic dispatch. Figure 1 shows the Synthesis architecture.

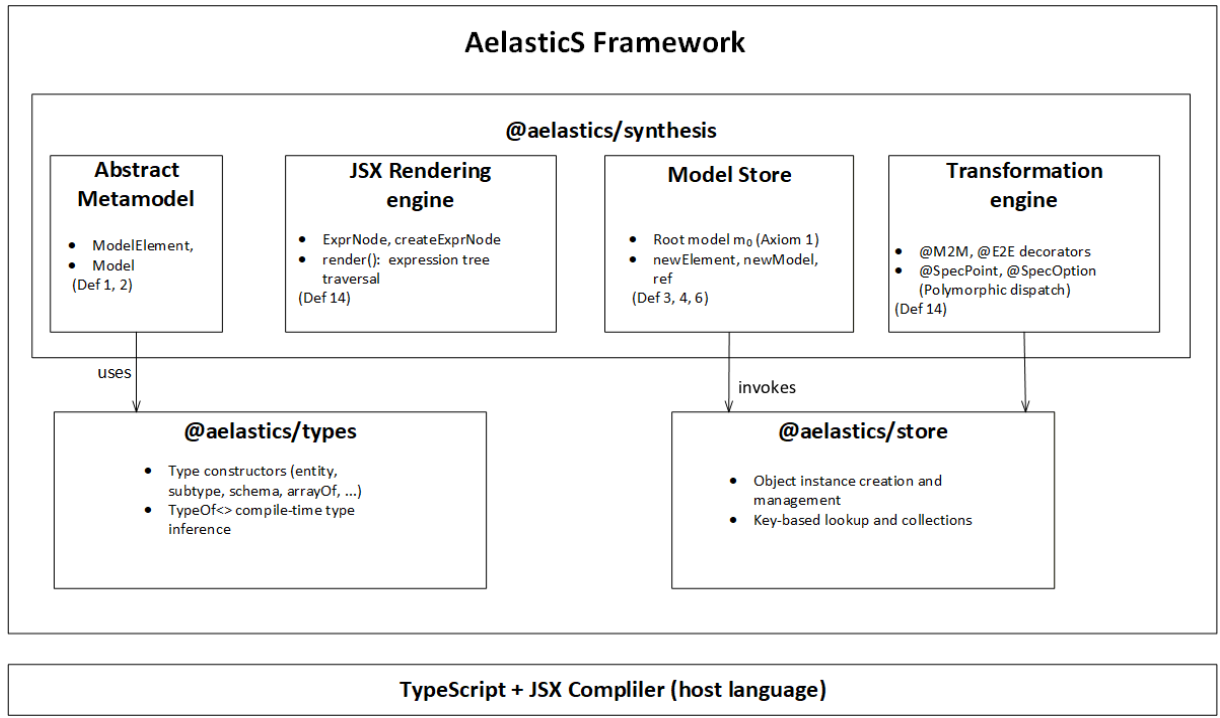

Figure 1. Synthesis architecture

The rest of the section is structured as follows. Section 4.1 presents JSX as the concrete syntax for model expressions. Section 4.2 shows how metamodels are defined as type schemas. Section 4.3 demonstrates model templates through parameterized and higher-order examples. Section 4.4 illustrates a complete model transformation.

### 4.1. JSX as Concrete Syntax for Model Expressions

JSX is an XML-like syntax extension to JavaScript, specified by Meta Platforms [12] independently of any particular framework. It is widely known through its use in React declarative construction of user interface component trees [13]. A JSX expression such as <Task name="Review"/> is compiled by the TypeScript compiler into a function call where the tag name, attributes, and nested children become arguments. Synthesis repurposes this compilation mechanism for model construction, with no dependency on React or any UI framework. By configuring the TypeScript compiler's jsxFactory option to redirect compilation to its own function instead of React's createElement function, Synthesis uses JSX as the concrete syntax for model expressions (Definition 13).

Each syntactic form in JSX corresponds to one of the four operators in a model expression. The model creation operator (μ) and element creation operator (ε) share the same JSX tag syntax, distinguished by whether the tag's type is a model type or an entity type:

(a) A JSX tag is a model creation operator (μ) when the tag name corresponds to a model type, and an element creation operator (ε) otherwise. The tag name determines the type, the attributes supply non-entity-typed property values, and the nested children supply entity-typed property values.

(b) The special $refByName attribute marks an element as a reference operator (ρ), invoking ref (Definition 6) to retrieve an existing element by name from the model under construction.

(c) Curly braces within JSX embed TypeScript expressions that produce child elements dynamically, corresponding to computation operators (κ). These include higher-order array functions such as map(), flatMap(), filter(), and reduce(), ternary expressions for conditional structure, and function calls for modular composition.

Listing 1 illustrates these forms in a single model expression.

```
const processName = "DocumentApproval"
const approvers = 2

export const Approval =
  <WorkflowModel name="SimpleApproval">
    <Process name="Publishing">
        <Sequence>
            <Task name="Format" />
            <Task name="Publish" />
        </Sequence>
    </Process>
    <Process name={processName}>
      <Sequence>
        <Task name="Write proposal" />
        <Parallel name="reviews">
          {Array.from({length: approvers}, (_, i) =>
            <Task name={`Approve ${i + 1}`} />
          )}
        </Parallel>
        <SubProcess $refByName="Publishing" />
      </Sequence>
    </Process>
  </WorkflowModel>
```

**Listing 1.** Model expression in JSX.

The variable Approval holds a model expression whose structure depends on the variables processName and `approvers` defined in the surrounding TypeScript scope. The WorkflowModel element serves as the container and namespace for all process definitions. The "Publishing" process is a model literal containing only element creation operators (ε). The second process is dynamic. The curly-brace expression {Array.from(...)} is a computation operator (κ) that generates a variable number of Task elements based on approverCount. The expression <SubProcess $refByName="Publishing" /> is a reference operator (ρ) that retrieves the already-created Publishing process by name and includes it as a step in the sequence.

The JSX tag names used in model expressions are not built-in. They are derived from the metamodel definition for the target domain, as shown in Section 4.2. The TypeScript type system ensures that only tags corresponding to valid entity types are accepted at compile time.

The compilation of these tags into executable code proceeds as follows. The TypeScript compiler translates each JSX tag into a call to the Synthesis createExprNode function (listing 2), configured via the jsxFactory compiler option.

```
export function createExprNode(
t: Template<IModelElement>, props: {},
...childrenExprs: ExprNode<any>[]): ExprNode<any> {
  let exprNode = t(props)
  exprNode.children.push(...childrenExprs.flat())
  return exprNode
}
```
Listing 2. The createExprNode function.

This function receives three arguments: an elementary template for the tag's type, a property record, and an array of child expression nodes. The elementary template is a function that creates an ExprNode for a specific entity type, recording a reference to its aelastics-types type descriptor and the name of the entity-typed property through which children are connected. The createExprNode function calls the elementary template with the given properties, attaches the child expression nodes, and returns the resulting ExprNode. Listing 3 shows elementary templates for two entity types from the Workflow metamodel.

```
...
export const Task: Template<ITask> = (props) => {
    return new ExprNode(Task, props, '')
}
export const Sequence: Template<ISequence> = (props) => {
    return new ExprNode(Sequence, props, 'steps')
}...
```
Listing 3. Elementary templates defining JSX tags.

The first argument to ExprNode is the type descriptor from the aelastics-types package. The second is the property record passed via JSX attributes. The third is the name of the entity-typed property that connects this element to its children in the model hierarchy. A Task has no entity-typed children, so the third argument is empty. A Sequence connects to its children through the 'steps' property. When a JSX expression such as <Sequence> <Task

name="Format"/> </Sequence> is compiled and executed, createExprNode calls the Sequence template to create a Sequence ExprNode, then attaches the Task ExprNode as a child under the 'steps' property. The result is a tree of ExprNode objects representing the model expression, ready for the second evaluation phase (Section 4.3).

### 4.2. Metamodel Definitions

Metamodels are defined using the type constructors realized as functions in the aelastics-types package. Each function produces a runtime type descriptor that records the type's structure, properties, and subtype relationships. The core constructors used in this paper are entity(), which produces identifiable types realizing c_ent, subtype(), which realizes the subtype constructor c_sub (Definition 8), schema(), which creates a type schema $\Sigma$ (Definition 11), and arrayOf(), optional(), and link(), which realize the array, optional, and reference constructors respectively.

A type schema $\Sigma = (T_\Sigma, \tau_{model})$ is created by calling schema(). Types are registered into a schema by passing the schema object as a parameter to each constructor call. Each constructor call produces a type descriptor object in memory. The TypeOf<> operator exploits TypeScript's conditional type inference to read that descriptor and derive a corresponding TypeScript interface at compile time. This means that a type definition written as a runtime object is automatically reflected into the static type system, with no code generation step.

Listing 4 defines the abstract metamodel that makes the type system model-aware by introducing two foundational types:

- ModelElement (Definition 1) is the base type for all model elements, providing each instance with a unique name.
- Model (Definition 2) is a special model element that serves as a container and namespace for a collection of other model elements.

Every concrete metamodel builds on these two types.

**Listing 4.** Abstract metamodel.
```
import { entity, subtype, schema, string, arrayOf } from "aelastics-types"

export const AbstractSchema = schema("Abstract-Schema")
export const ModelElement = entity({
    name: string,
}, ["name"], "ModelElement", AbstractSchema)

export const Model = subtype(ModelElement, {
    elements: arrayOf(ModelElement),
}, "Model", AbstractSchema)

export type IModelElement = TypeOf<typeof ModelElement>
export type IModel = TypeOf<typeof Model>
```

The second argument to entity() specifies which properties serve as the identifying key of he created instance. Because the aelastics-types package is a general-purpose type system unaware of models, uniqueness of element names within a model is thus enforced by the Synthesis runtime during evaluation (Definition 3).

Listing 5 defines the Workflow metamodel, which serves as the target type schema $\Sigma_T$ for the transformation in Section 4.4.

**Listing 5.** Workflow metamodel.

```
import { subtype, schema, arrayOf, optional, string } from "aelastics-types"
import { ModelElement, Model } from "aelastics-synthesis"

export const WF_Schema = schema("Workflow-Schema")

export const Step = subtype(ModelElement, {}, "Step", WF_Schema)

export const Task = subtype(Step, {
    performer: optional(string),
}, "Task", WF_Schema)

export const Sequence = subtype(Step, {
    steps: arrayOf(Step),
}, "Sequence", WF_Schema)

export const Parallel = subtype(Step, {
    steps: arrayOf(Step),
}, "Parallel", WF_Schema)

export const Process = subtype(Step, {
    flow: Step,
}, "Process", WF_Schema)

export const WorkflowModel = subtype(Model, {
    processes: arrayOf(Process),
}, "WorkflowModel", WF_Schema)

export type IStep = TypeOf<typeof Step>
export type ITask = TypeOf<typeof Task>
export type IProcess = TypeOf<typeof Process>
export type IWorkflowModel = TypeOf<typeof WorkflowModel>
```

A WorkflowModel subtypes Model and serves as the container for process definitions. Each Process consists of a top-level Step, specified by the flow property. A step can be a Task with an optional performer, a Sequence of steps, a Parallel composition of steps, or a sub-process. Because Process itself subtypes Step, a process can appear as a step within another process, enabling hierarchical composition as illustrated by the $refByName example in Listing 1. The JSX tag names used in Listing 1 are derived from these type definitions.

### 4.3. Model Expressions, Templates, and Evaluation

The formal framework distinguishes model literals (ground terms containing only μ, ε, and ρ operators), model templates (expressions with computation operators κ referencing parameters from a parameter space Θ, Definition 15), and higher-order templates (templates where at least one parameter is itself a template, Definition 16). In TypeScript, these distinctions are expressed through function boundaries. A model literal is a JSX expression with all variables bound in the surrounding scope, producing a fixed expression tree immediately. A model template is a function whose parameters constitute $\Theta$, and the JSX expression in the function body cannot be evaluated until the function is called with concrete arguments. A higher-order template is realized through currying. The first function call binds the template parameter in the closure, returning a regular template that can be reused in model expressions. The second call during expression evaluation binds the data parameters and produces the expression tree.

Evaluation of a model expression involves a compile-time step and two runtime phases (Figure 2). At compile time, the TypeScript compiler translates each JSX tag into a call to createExprNode (Section 4.1). In the first runtime phase, the JavaScript interpreter executes these calls, producing a tree of ExprNode objects that represent the model expression by recording type descriptors, properties, and children without yet creating model elements in the Model Store. Embedded TypeScript expressions within curly braces are also evaluated during this phase, reducing $\kappa$ nodes to concrete sub-expressions. For a model literal, this phase executes automatically. For a template, which is a function definition, this phase executes when the template function is called with concrete arguments.

In the second runtime phase, store.render() passes the top-level ExprNode to the JSX Rendering Engine (Figure 1), together with the Model Store as the enclosing model. The Model Store plays the role of the root model $m_0$ (Axiom 1), providing the outermost namespace and universal type schema. The JSX Rendering Engine performs a recursive traversal of the expression tree in document order (Definition 14), instantiating elements in the Model Store via newElement (Definition 3) for each $\varepsilon$ node and resolving references via ref (Definition 6) for each $\rho$ node. When the traversal encounters a $\mu$ node, newModel creates a new model within the current enclosing model, and its child expressions are evaluated recursively within that new model. A $\mu$ node nested inside another $\mu$ node produces a megamodel (Section 3.4.3). A $\rho$ node that resolves to an element in a different model already present in the Model Store produces a cross-model reference, enabling weaving models (Section 3.4.4).

Figure 2 illustrates these steps, from JSX compilation through Element expression tree construction to model instantiation in the store.

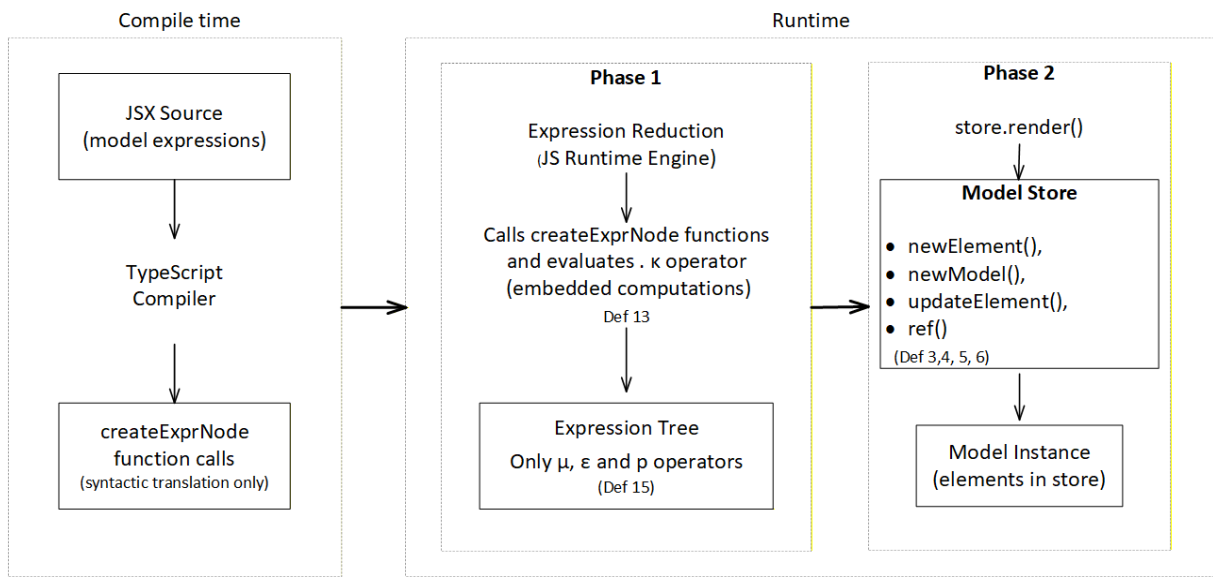

Figure 2. Evaluation phases of a model expression

The following listings illustrate how model templates and higher-order templates are used in practice. Listing 6 defines ConfigurableApproval, a model template that generates approval workflows varying in document name, number of approvers, and structural arrangement.

**Listing 6.** ConfigurableApproval.

```
type IApprovalConfig = {
  document: string
  approvers: number
  mode?: "parallel" | "sequential"
}

const ConfigurableApproval = ({document, approvers, mode}:IApprovalConfig) =>
{
  const tasks = new Array(approvers)
    .map((_, i) => <Task name={`${document}Approval-${i}`} />)
  return (
    <WorkflowModel name={`ApprovalWF-${document}`}>
      <Process name={`Approve ${document}`}>
        <Sequence>
          <Task name={`Write ${document}`} />
          {mode === "parallel"
            ? <Parallel> {tasks} </Parallel>
            : <Sequence> {tasks} </Sequence>}
        </Sequence>
      </Process>
    </WorkflowModel>)
}

// usage
const ContractApproval = <ConfigurableApproval
  document="Contract"
  approvers={3}
  mode="parallel"
/>
```

```
// evaluate ContractApproval expression to get model value
const store = new ModelStore()
const contractApprovalModel = store.render(ContractApproval)
```

The IApprovalConfig type defines the parameter space $\Theta$. The Array(approvers).map(…) expression is a computation operator ($\kappa$) that performs iterative generation, producing as many Task elements as there are approvers. The ternary expression is another $\kappa$ that selects between a Parallel and a Sequence arrangement. The JSX invocation <ConfigurableApproval document="Contract" approvers={3} mode="parallel" /> calls the template function with concrete arguments, producing a model expression. Calling store.render() on that expression triggers the second runtime phase, instantiating the model and its elements in the store.

The result is an IWorkflowModel object, a fully populated model value with the following structure:

```
WorkflowModel "ApprovalWF-Contract"
└── Process "Approve Contract"
    └── Sequence
        ├── Task "Write Contract"
        └── Parallel
            ├── Task "ContractApproval-0"
            ├── Task "ContractApproval-1"
            └── Task "ContractApproval-2"
```

With {document: "Invoice", approvers: 2, mode: "sequential"}, the same template produces a structurally different model with two sequential approval tasks. The two resulting models differ in cardinality, topology, and element properties. JSX requires that a template function receive its parameters as a single props object, so the IApprovalConfig type serves both as the formal parameter space $\Theta$ and as the required props type for JSX invocation.

Listing 7 defines GenericApproval, a higher-order template (Definition 16) that accepts another template as a parameter.

**Listing 7.** GenericApproval: higher-order template.

```
const GenericApproval = (WorkerTask: Template<ITask>) =>
    (c: IApprovalConfig) => {
        const tasks = new Array(c.approvers)
            .map((_, i) => <Task name={`${c.document}Approval-${i}`} />)
        return (
          <WorkflowModel name={`ApprovalWF-${c.document}`}>
            <Process name={`Approve ${c.document}`}>
              <Sequence>
                <WorkerTask name={`Write ${c.document}`} />
                {c.mode === "parallel"
                  ? <Parallel> {tasks} </Parallel>
                  : <Sequence> {tasks} </Sequence>}
              </Sequence>
            </Process>
          </WorkflowModel>
        )
    }

const TwoStepWrite: Template<ITask> = ({ name }) => (
```

```
        <Sequence>
            <Task name={`${name}-draft`} />
            <Task name={`${name}-final`} />
        </Sequence>
)

const TwoStepApproval = GenericApproval(TwoStepWrite)
const contractTwoStep = (
  <TwoStepApproval document="Contract" approvers={2} mode="parallel" />
)
const contractTwoStepModel = store.evaluate(contractTwoStep)
```

The parameter WorkerTask: Template<ITask> is a template-typed parameter, a function that produces a model expression returning a Task model element when evaluated. The first call, GenericApproval(TwoStepWrite), binds it in the closure and returns a regular template TwoStepApproval. The second call, via JSX invocation, binds the data parameters and produces the expression tree. After evaluation, the model value has the following structure:

```
WorkflowModel "ApprovalWF-Contract"
 └─ Process "Approve Contract"
     └─ Sequence
         ├─ Sequence                                     ← TwoStepWrite expanded
         │   ├─ Task "Write Contract-draft"
         │   └─ Task "Write Contract-final"
         └─ Parallel
             ├─ Task "ContractApproval-0"
             └─ Task "ContractApproval-1"
```

TwoStepWrite replaces a single Task with a two-step draft-and-final sequence. A different template passed to GenericApproval would produce a different initial work activity while sharing the same approval structure. This directly realizes Definition 16, where the κ operator within the template invokes a template-typed parameter to produce sub-expressions.

Table 1 summarizes how the formal model expression concepts map to TypeScript and JSX mechanisms.

**Table 1.** Model expression concepts mapped to TypeScript/JSX.

| Formal concept | TypeScript realization | Runtime result |
| --- | --- | --- |
| Model expression (Def. 13) | JSX expression | Element expression |
| Model literal (ground term) | JSX with all variables bound in scope | Element tree expression, produced immediately |
| Model template (Def. 15) | Function returning JSX | Element tree expression, produced on call |
| Higher-order template (Def. 16) | Curried function with template parameter | Template, then Element tree expression |
| Model value | Result of store.render() | Model elements (objects) in the store |

### 4.4. Model Transformations

The model expression algebra formalized in Section 3 covers not only model construction and templates (Sections 3.1–3.4) but also transformation specifications with typed rules, polymorphic dispatch, and execution semantics (Sections 3.5–3.6). Realizing transformation specifications in a programming language requires mechanisms that go beyond what model expression evaluation alone provides. Synthesis realizes these requirements through a rule-based transformation framework. A transformation specification (Definition 18) is realized as a TypeScript class extending a predefined abstract transformation base class, and transformation rules are methods whose bodies are model templates producing target model expressions.

TypeScript classes and methods provide the natural structure for rules, but the metadata that connects them has no direct syntactic counterpart in the language. Hence, we have decided to use TypeScript decorators to bridge this gap. Decorators are a metaprogramming mechanism in TypeScript that provides aspect-oriented capabilities. The framework defines two categories of decorators: trace decorators and dispatch decorators.

Trace decorators record source-to-target mappings for the execution trace (Definition 22): @M2M() on the class registers the transformation itself, linking the source and target type schemas. @E2E() on a method registers it as a transformation rule (Definition 17), enabling automatic trace generation for each invocation.

Dispatch decorators implement polymorphic rule dispatch (Definition 19): @SpecPoint() marks a method as a polymorphic dispatch target. @SpecOption(pointName, sourceType) registers a method as a specialization of the named dispatch point for a specific source subtype.

### 4.4.1. Source and Target Metamodels

The target metamodel (Workflow) was defined in Listing 5 (Section 4.2). Listing 8 presents the OrgModel metamodel, the source type schema $\Sigma_S$.

**Listing 8.** OrgModel metamodel (source).

```
import { subtype, schema, arrayOf, optional, string, link } from "aelastics-types"
import { ModelElement, Model } from "aelastics-synthesis"

export const Org_Schema = schema("OrgModel-Schema")

export const Worker = subtype(ModelElement, {
    role: optional(string),
    worksIn: optional(link(Org_Schema, "OrgUnit")),
}, "Worker", Org_Schema)

export const OrgUnit = subtype(ModelElement, {
    parent: optional(link(Org_Schema, "OrgUnit")),
    subUnits: arrayOf(link(Org_Schema, "OrgUnit")),
    members: arrayOf(Worker),
}, "OrgUnit", Org_Schema)

export const Department = subtype(OrgUnit, {
    manager: Worker,
}, "Department", Org_Schema)
```

```
    export const Board = subtype(OrgUnit, {
    }, "Board", Org_Schema)

    export const Organization = subtype(Model, {
        topUnit: OrgUnit,
    }, "Organization", Org_Schema)

    inverseProps(OrgUnit, "parent", OrgUnit, "subUnits")
    inverseProps(OrgUnit, "members", Worker, "worksIn")

    export type IWorker = TypeOf<typeof Worker>
    export type IOrgUnit = TypeOf<typeof OrgUnit>
    export type IDepartment = TypeOf<typeof Department>
    export type IBoard = TypeOf<typeof Board>
    export type IOrganization = TypeOf<typeof Organization>
```

Organization subtypes Model and designates a topUnit as the root of the hierarchy. Worker represents an individual with an optional role and a worksIn link. OrgUnit is the common supertype, carrying parent and subUnits links that form the hierarchy, and a members array. Department adds a manager reference. Board adds no properties. The inverseProps calls establish bidirectional relationships.

### 4.4.2. Transformation Specification

Listing 9 presents the source model instance.

**Listing 9.** OrgModel source model instance.

```
    export const AcmeCorp =
      <Organization name="AcmeCorp" MDA_level="M1">
        <Board name="Executive Board">
          <Worker name="Alice" role="Chair" />
          <Worker name="Bob" />
          <Worker name="Carol" />
        </Board>
        <Department name="Technology Division"
          parent={<OrgUnit $refByName="Executive Board" />}>
          <Worker name="Dave" role="VP" />
        </Department>
        <Department name="Engineering"
          parent={<OrgUnit $refByName="Technology Division" />}>
          <Worker name="Grace" role="Director" />
          <Worker name="Judy" />
        </Department>
      </Organization>
```

Listing 10 presents the transformation specification, parameterized by ITransformParam.

**Listing 10.** OrgModel-to-ApprovalWorkflow transformation.

```
    /** @jsx hm */
    import { hm } from "aelastics-synthesis"
    import { SpecPoint, SpecOption } from "aelastics-synthesis"
    import { abstractM2M, M2M, E2E } from "aelastics-synthesis"
    import { Element } from "aelastics-synthesis"
```

```typescript
import * as ot from "./org-model.meta"
import * as wt from "./workflow.meta"
import * as w from "./workflow.jsx-comps"

export interface ITransformParam {
  worker: ot.IWorker
  sensitivity: number
}

@M2M()
export class Org2WorkflowTransformation extends abstractM2M<
  ot.IOrganization,
  wt.IWorkflowModel,
  {},
  never,
  ITransformParam
> {
  template(source: ot.IOrganization) {
    const { worker, sensitivity } = this.context.param!
    const unfold = (unit: ot.IOrgUnit, levels: number):
        ot.IOrgUnit[] =>
      (levels <= 0) ? []
        : unit.parent
          ? [unit, ...unfold(unit.parent as ot.IOrgUnit,
              levels - 1)]
          : [unit]
    return (
      <w.WorkflowModel name="Approval" MDA_level="M1">
        <w.Process name={`ApprovalFor${worker.name}`}>
          <w.Flow name="ApprovalSequence">
            <w.Task name="WriteDocument"
              performer={worker.name} />
            {unfold(worker.worksIn as ot.IDepartment,
                sensitivity)
              .map(unit =>
                this.OrgUnit2Approval(unit))}
          </w.Flow>
        </w.Process>
      </w.WorkflowModel>
    )
  }

  @E2E()
  @SpecPoint()
  OrgUnit2Approval(unit: ot.IOrgUnit): Element<wt.IStep> {
    return (
      <w.Task name={`ApproveBy${unit.name}`} />
    )
  }

  @SpecOption("OrgUnit2Approval", ot.Department)
  Department2Approval(dept: ot.IDepartment):
      Element<wt.ITask> {
    return (
      <w.Task name={`ApproveBy${dept.manager.name}`}
        performer={dept.manager.name} />
    )
  }
```

```
    @SpecOption("OrgUnit2Approval", ot.Board)
    Board2Approval(board: ot.IBoard):
        Element<wt.IParallel> {
      return (
        <w.Parallel name={`BoardApproval${board.name}`}>
          {board.members.map(member =>
            <w.Task name={`ApproveBy${member.name}`}
              performer={member.name} />
          )}
        </w.Parallel>
      )
    }
  }
```

The template() method is the top-level rule r₀. It receives the source Organization model and accesses the transformation parameters from the context. The unfold function walks up the hierarchy from the worker's department through parent links, collecting at most sensitivity organizational units. The method returns a JSX model expression that constructs a WorkflowModel containing a single Process. The .map() expression is a computation operator (κ) that invokes the dispatch mechanism for each unit in the chain.

OrgUnit2Approval is decorated with @E2E() and @SpecPoint(), making it both a traced rule and a polymorphic dispatch target. Two specialization options handle the structurally different cases. Department2Approval produces a single Task assigned to the department's manager. Board2Approval produces a Parallel element containing one Task per board member. No conditional branching appears in the transformation code, as the dispatch is automatic.

### 4.4.3. *Execution and Traceability*

Consider invoking the transformation for worker Judy with sensitivity 3:

```
const transformation = new Org2WorkflowTransformation()
const result = transformation.transform(AcmeCorp,
    { worker: judy, sensitivity: 3 })
```

The unfold function starts from Judy's department "Engineering" and walks up, collecting three units: (1) "Engineering" (Department), (2) "Technology Division" (Department), (3) "Executive Board" (Board). For "Engineering," dispatch selects Department2Approval, producing a task assigned to Grace. For "Technology Division," dispatch again selects Department2Approval, producing a task assigned to Dave. For "Executive Board," dispatch selects Board2Approval, producing three parallel tasks assigned to Alice, Bob, and Carol. Listing 11 shows generated target model.

**Listing 11.** Generated target model.

```
<WorkflowModel name="Approval">
  <Process name="ApprovalForJudy">
    <Sequence name="ApprovalSequence">
      <Task name="WriteDocument"
        performer="Judy" />
      <Task name="ApproveByGrace"
        performer="Grace" />
```

```
        <Task name="ApproveByDave"
          performer="Dave" />
        <Parallel name="BoardApprovalExecutive Board">
          <Task name="ApproveByAlice"
            performer="Alice" />
          <Task name="ApproveByBob"
            performer="Bob" />
          <Task name="ApproveByCarol"
            performer="Carol" />
        </Parallel>
      </Sequence>
    </Process>
  </WorkflowModel>
```

The framework automatically constructs an execution trace recording the correspondence between source and target elements. Listing 12 shows representative trace entries.

**Listing 12.** Execution trace (representative entries).

```
<TraceModel name="AcmeCorp to Approval" timestamp="2026-04-02T00:12:00.173Z"
source={<Model $refByName="/AcmeCorp" />} targets={[<Model $refByName="/Approval"
/>]}>
    <TraceEntry source={<ModelElement $refByName="/AcmeCorp/Engineering" />}
    rule="OrgUnit2Approval" ruleType="RegularRule" targets={[<TraceEntry
    $refByName="/Approval/ApproveByGrace" />]} />

    <TraceEntry source={<ModelElement $refByName="/AcmeCorp/TechnologyDivision"
    />} rule="OrgUnit2Approval" ruleType="RegularRule" targets={[<TraceEntry
    $refByName="/Approval/ApproveByDave" />]} />

    <TraceEntry source={<ModelElement $refByName="/AcmeCorp/ExecutiveBoard" />}
    rule="OrgUnit2Approval" ruleType="RegularRule" targets={[<TraceEntry
    $refByName="/Approval/BoardApprovalExecutiveBoard" />]} />
</TraceModel>
```

Each trace entry records the source element, the produced target elements, and the dispatched rule, forming the triple ($e_k$, $r_k$, $t_k$) of Definition 22. The trace satisfies Proposition 4. Dispatch determinism (Proposition 3) is satisfied because Department and Board are distinct subtypes of OrgUnit with no overlap. Dispatch completeness (Definition 20) holds because every OrgUnit subtype has an applicable rule. Type safety (Theorem 2) is maintained because every rule produces well-typed target model expressions over the Workflow type schema.

This completes the realization of the formal framework in the Synthesis framework. Table 2 summarizes how each concept from the formal framework maps to a concrete mechanism in Synthesis.

**Table 2.** Mapping of transformation concepts to Synthesis mechanisms.

| Formal Concept | Synthesis Realization |
|---|---|
| Transformation specification (Def. 18) | Class decorated with @M2M |
| Top-level rule $r_0$ | template() method |
| Regular rule (Def. 17) | Method decorated with @E2E |

| Specialization point | Method decorated with @SpecPoint and @E2E |
| Specialization option | Method decorated with @SpecOption |
| Rule dispatch (Def. 19) | Runtime type inspection via store |
| Execution trace (Def. 22) | Automatically recorded by decorator infrastructure |
| Transformation parameters | this.context.param via generic type P |

## 5. Related Work

This section positions our functional approach against existing work on model representation and transformation.

### 5.1. Model Driven Architecture and Frameworks

The OMG Model Driven Architecture (MDA) [3] and the Eclipse Modeling Framework (EMF) [4] constitute the reference infrastructure for model-driven development. MOF [7] provides the metamodeling layer, QVT [7] provides transformation languages (declarative QVTr, imperative QVTo, and the QVT Core kernel), and OCL provides constraint specification. EMF operationalizes this vision through ECore, a practical subset of MOF, and a rich ecosystem of tools. The Atlas Transformation Language (ATL) [8], the most widely adopted transformation language in the EMF ecosystem, provides a hybrid declarative-imperative paradigm with matched rules, called rules, and imperative blocks.

Höppner et al. [14] provide an empirical perspective on the relationship between dedicated transformation languages and general-purpose languages. Their longitudinal study finds that modern Java (SE14) reduced transformation complexity by approximately 45% compared to older Java versions, yet ATL still hides approximately 40% of the complexity that remains visible in Java. These findings support the observation that mainstream languages are closing the gap with dedicated transformation languages. Our approach combines a mainstream language (TypeScript) with algebraic foundations that provide formal guarantees absent from both ATL and plain Java.

The fundamental difference between MDA/EMF and our approach lies in model representation. In MDA/EMF, models are object graphs, networks of interconnected objects instantiated from metamodel classes. A model is a passive data structure. In our approach, models are values produced by evaluating model expressions, term-algebraic compositions of $\mu$, $\varepsilon$, $\rho$, and $\kappa$ operators (Section 3.4). This has several consequences. Parameterization is immediate, since a model expression with free variables is a template defining an entire family of models. Transformations require no separate language, since they are templates over source models (Section 3.5) expressed in the same language as models and metamodels. Three of four conformance conditions are enforced statically at compile time (Proposition 1). The model expression algebra provides compositional semantics that the ad hoc composition mechanisms of MDA/EMF lack.

Two further consequences concern megamodels and weaving models. In the MDA/EMF ecosystem, megamodels [10], [11] and weaving models [12] [13] are separate architectural concepts requiring dedicated tooling. Aničić et al. [14] showed that the original weaving metamodels further require external OCL constraints to enforce valid mapping rules. In our algebra, megamodels arise because models are model elements (Definition 2), and weaving models arise because c_iref properties (Definition 8) encode typed cross-model references, with conformance checking (Definition 12) replacing external constraints (Sections 3.4.3, 3.4.4)

We acknowledge that MDA and EMF provide a mature ecosystem with extensive industrial adoption, rich tooling, and a large body of existing metamodels and transformations. However, the limitations we identify, including passive model representation, separate transformation languages, limited parameterization, and fragmented formal foundations, are inherent to the object-oriented paradigm rather than incidental to any particular implementation.

### 5.2. Graph-Based Approaches

Graph-based transformation approaches formalize models as typed attributed graphs and transformations as graph rewriting rules. Triple Graph Grammars (TGGs), introduced by Schürr [15], specify transformations through grammar rules that simultaneously describe source patterns, target patterns, and their correspondence, providing native support for bidirectional transformations. Algebraic graph transformation frameworks, including AGG [16], Henshin [17], and the theoretical foundations of Ehrig et al. [18], formalize graph transformations through categorical pushout constructions.

Graph-based approaches share with our work a commitment to rigorous formalization, but the formalisms differ in a way that has practical consequences. Graph-based approaches formalize transformations as graph morphisms or rewriting rules, requiring dedicated tools with specialized execution engines. Our approach formalizes models as values in a term algebra and transformations as templates over source models, using mathematical structures that map directly to constructs in a mainstream programming language (Section 4). Type safety in graph-based approaches is verified by the tool against a type graph. In our approach, it is enforced by the host language's type checker at compile time (Proposition 1).

TGGs provide native bidirectionality, which our framework does not address. This is a direction for future work (Section 6), potentially through the lens formalism discussed in Section 5.3. Conversely, TGGs and algebraic graph transformations lack natural parameterization and template mechanisms. A TGG rule defines a fixed structural correspondence, with no direct mechanism for generating different correspondence patterns based on parameters.

An important distinction concerns how each approach uses algebraic methods. Algebraic graph transformation uses algebra to describe graph rewriting, the controlled replacement of subgraphs within a host graph, through categorical vocabulary (pushouts, graph morphisms). Our approach uses algebra to describe model construction, where model expressions are terms and evaluation is the interpretation homomorphism from terms to values, through classical vocabulary (sets, functions, type constructors). These are complementary uses of algebraic formalism.

### 5.3. Functional and Bidirectional Approaches

The lens framework, originating from Foster et al. [19] and refined by Hofmann et al. [20], formalizes bidirectional transformations as pairs of get and put functions satisfying round-tripping laws. Diskin et al. [21] generalize lenses to multiary settings for model synchronization. Our framework is complementary. The models and transformations we formalize could potentially be organized into lens structures to achieve bidirectionality, and the compositional nature of the model expression algebra aligns well with lens combinators. We leave this integration as future work.

Rensink [22] proposed a functional approach to graph-based model transformation, representing transformation rules as functions over graph structures, but the approach remains within the graph-based formalism and does not extend to model representation or metamodeling. The Stratego/XT framework [23] provides rewriting strategies for program transformation, treating programs as terms and transformations as term-rewriting rules composed through strategy combinators. Two differences from our approach are significant. Stratego operates on untyped syntax trees rather than on typed model elements conforming to a metamodel. There is no notion of a model store providing named element registration and cross-reference resolution, which is essential for representing graph-structured models within a term algebra. The Spoofax language workbench [24] extends the Stratego tradition to language engineering but focuses on textual language processing rather than model-driven engineering.

Diskin [25] proposed a categorical framework for model management, and Rutle et al. [26] used category theory to formalize metamodel-based transformations through diagram predicates. Our formal framework (Section 3) uses standard mathematical structures (sets, functions, partial orders) rather than categorical language, keeping the formalization accessible. The category-theoretic perspective, particularly monadic composition of transformations, is a direction for future work (Section 6).

### 5.4. Algebraic and Type-Theoretic Foundations

Our model expression algebra instantiates the classical pattern of initial algebra semantics, pioneered by the ADJ group [27] and developed further by Goguen and Meseguer [28]. In that tradition, a signature defines sorts and operations, the term algebra over the signature is the initial algebra, and for any algebra there exists a unique homomorphism from terms to values. Our signature consists of model creation operators $\mu$, element creation operators $\varepsilon$, reference operators $\rho$, and computation operators $\kappa$. Model expressions are terms, models are values, and the evaluation function (Definition 14) is the interpretation homomorphism. The key difference from classical algebraic specification is that our algebra is parameterized by a type schema $\Sigma$ that defines the available types, their properties, and their relationships. The algebra itself varies with the metamodel, reflecting the MDE principle that the metamodel governs the space of valid models.

The embedded DSL tradition, articulated by Hudak [29] and extended through the tagless final style of Carette et al. [30], provides a related pattern where DSL terms are interpreted by different back-ends. Our approach shares the EDSL philosophy, with JSX as concrete syntax and TypeScript as host language. However, our setting introduces domain-specific structure absent from general-purpose EDSLs, including metamodel-parameterized signatures, a model store providing named element registration and cross-reference resolution, and relationship-directionality handling.

The relationship between type systems and metamodeling has been explored from several perspectives. De Lara and Guerra [31] developed multilevel metamodeling beyond the standard four-level architecture, with potency-based typing. Atkinson and Kühne [32] proposed deep instantiation for defining properties at arbitrary metalevels. Guy et al. [33] explored how type theory can encode metamodel conformance, and Steel and Jézéquel [34] proposed parameterized metaclasses with type arguments. Our framework realizes the M3→M2→M1 architecture through the host language's type system rather than designing a specialized type-theoretic framework. We do not support multilevel modeling or deep instantiation. The TypeOf<> mechanism (Section 4.2) bridges runtime type descriptors and compile-time types, providing practical IDE support that specialized frameworks typically lack. The subtype relation drives polymorphic dispatch (Section 3.5), connecting the metamodel's type structure directly to transformation semantics, a connection absent in approaches treating type systems and transformation languages as separate concerns.

### 5.5. Template and Reuse Mechanisms in MDE

Kusel et al. [9] provide the most comprehensive survey of reuse in model transformation languages, evaluating 13 reuse mechanisms across five transformation scenarios. Their central finding is that no single mechanism fully addresses reuse across all scenarios, and reuse across metamodel boundaries remains an unsolved problem. The survey won the 2025 MODELS ten-year most influential paper award, confirming the continued relevance of this challenge.

Existing reuse mechanisms fall into three categories. UML templates [35] allow classifiers and packages to be parameterized by type parameters but are restricted to type substitution and cannot parameterize model structure such as element cardinality, conditional inclusion, or computed properties. ATL rule inheritance [8] allows transformation rules to extend other rules, inheriting matching conditions and creation logic, but is confined to the ATL ecosystem and applies only to transformation rules, not to model construction. Package templates, proposed by Krogdahl et al. [36], support additive composition of model fragments through renaming and merging, but do not support value-dependent structural variation.

In our framework, the template mechanism arises directly from the model expression algebra (Section 3.4). This provides three levels of reuse along an expressiveness spectrum. Regular templates provide value parameterization, where parameters control element properties, cardinality, and structural alternatives. Higher-order templates provide template parameterization, where a template accepts another template as argument, enabling patterns such as a generic approval process that accepts an arbitrary work activity template (Section 4.3). A third level, schema-polymorphic templates, would extend reuse across metamodel boundaries. The type system already supports accepting elements from different metamodels through generic type parameters (e.g., Template<T extends IModelElement>), but full schema-polymorphic templates require parameterization of template inputs and outputs by type, which JSX syntax does not currently support. This remains future work. The first two levels already address a significant portion of the gap identified by Kusel et al., and the third level targets the cross-metamodel reuse that their survey identifies as the central unsolved problem.

## 6. Conclusion and Future Work

### 6.1. Summary of Contributions

This paper has shown that model construction and model transformation can be unified within a single algebraic framework. The model expression algebra, built on four operators, model creation ($\mu$), element creation ($\varepsilon$), reference ($\rho$), and computation ($\kappa$), provides compositional semantics for assembling models as values from hierarchically nested expressions. Metamodels, formalized as type schemas, constrain the algebra by specifying which constructors exist and what their argument types must be. Templates arise naturally as parameterized model expressions, open terms with free variables that define entire families of models from a single definition. Transformations are then simply templates whose parameter space consists of source model elements, requiring no separate language or formalism. The formal framework establishes two central results: type preservation under evaluation (Theorem 1), guaranteeing that well-typed expressions produce conforming models, and type safety of transformation execution (Theorem 2), guaranteeing that well-typed rules produce conforming target models. The Synthesis framework validates these ideas through an embedded DSL in TypeScript and JSX, where three of four conformance conditions are enforced statically by the host language's type system and a single mainstream language serves simultaneously as the metamodeling, model construction, and transformation language.

### 6.2. Limitations

The framework currently supports only unidirectional transformations. Bidirectional model transformations, essential for model synchronization and round-trip engineering, require substantial extensions to the formalization, potentially through lens-based approaches. Incremental transformations are not yet addressed. The explicit trace structures and compositional model expressions provide a foundation for dependency tracking, but mechanisms for incremental re-execution remain to be formalized. The tooling ecosystem surrounding Synthesis is less mature than MDA/EMF, which provides graphical editors, model repositories, and interchange formats developed over two decades. Empirical validation in this paper is limited to illustrative examples. The open-source Synthesis framework includes additional examples, but large-scale industrial case studies and controlled experiments comparing functional MDD with OO-based approaches remain future work.

### 6.3. Future Research Directions

The formal foundations established in this paper open several research directions, ranging from expressiveness analysis through algebraic extensions to practical unification of transformation paradigms and integration with generative AI.

We believe that functional templates can express a wider range of model parameterization, structural variation, and transformation logic reuse than existing MDE reuse mechanisms. The comprehensive evaluation framework of Kusel et al. [9], which assesses reuse mechanisms across several transformation scenarios, provides a natural basis for evaluating this claim. A rigorous comparison remains future work.

The algebraic formalization established in this paper could be enriched with category-theoretic concepts to provide additional structure for reasoning about models and transformations. One specific direction is to investigate whether models can be viewed as monadic values and templates as monadic functions. If so, sequential transformation composition could be formalized as monadic composition, with the monad laws providing algebraic guarantees about transformation pipelines.

A related direction is bidirectional model transformations through the lens formalism (Section 5.3). The compositional structure of model expressions aligns well with lens combinators, and investigating whether model templates can be organized into lens pairs satisfying round-tripping laws would extend the framework to model synchronization and round-trip engineering.

A further direction concerns design decisions in transformations. In practice, choices such as whether a department uses sequential or parallel approval are hard-coded in transformation rules. Externalizing such variability into separate, configurable decision models would allow the same transformation logic to produce different target structures based on configuration. We address this direction in [37], where we formalize configurable decision models and provide an implementation within the Synthesis framework.

Current MDD practice treats model-to-model and model-to-text transformations as fundamentally different activities requiring separate languages and tools. If model-to-text transformation can be treated as a special case of model-to-model transformation, through a predefined textual content metamodel whose instances represent files, directories, and textual elements, then the same rule dispatch, trace generation, and type safety properties established in this paper would apply equally to code generation and documentation generation.

The rise of generative AI and large language models opens a further research direction. Because model expressions, templates, and transformations are expressed as typed TypeScript and JSX code, this creates the possibility of AI-assisted generation of model expressions and transformation rules. On the other hand, our framework's formal structure with its expressiveness and flexibility can guide and constrain AI-genAnierated model artifacts and code. Investigating how generative AI and model-driven engineering based on our functional approach can synergistically merge is a promising direction for future work.

### 6.4. Closing Statement

The functional paradigm offers a formally grounded alternative to the object-oriented approaches that have shaped model-driven engineering for over two decades. By treating models as values in a typed algebra and transformations as templates over source models, the approach provides compositional semantics, static type safety, and parameterization capabilities that are difficult to achieve within the object-oriented paradigm. At the same time, models, metamodels, and transformations can be mapped to ordinary artifacts within a mainstream programming ecosystem, using existing tools and offering a path toward broader adoption of model-driven engineering in the software industry.